\documentstyle[12pt,fleqn,cite,epsfig,psfrag]{article}

\def\be{\begin{equation}}
\def\ee{\end{equation}}
\def\bea{\begin{eqnarray}}
\def\eea{\end{eqnarray}}

\newcommand{\beqa}{\begin{eqnarray}}
\newcommand{\eeqa}{\end{eqnarray}}
\newcommand {\tr}{\mbox{tr}}
\newcommand {\expt}[1]{\left\langle #1 \right\rangle}

\begin{document}
\baselineskip=0.6cm
\begin{titlepage}
\begin{center}
\hfill hep-th/0609038\\
\hfill IP/BBSR/2006-17\\
\vskip .2in

{\Large \bf Phase Transitions in Higher Derivative Gravity}
\vskip .5in

{\bf Tanay K. Dey\footnote{e-mail: tanay@iopb.res.in}, Sudipta
Mukherji\footnote{e-mail: mukherji@iopb.res.in}, Subir Mukhopadhyay
\footnote{e-mail: subir@iopb.res.in} and Swarnendu Sarkar
\footnote{e-mail: swarnen@iopb.res.in}\\
\vskip .1in
{\em Institute of Physics,\\
Bhubaneswar 751 005, INDIA}}

\end{center}

\begin{center} {\bf ABSTRACT}
\end{center}
\begin{quotation}\noindent
\baselineskip 15pt

This paper deals with black holes, bubbles and orbifolds in Gauss-Bonnet theory in five dimensional anti de 
Sitter space.  In particular, we study stable, unstable and metastable phases of black holes
from thermodynamical perspective. By comparing bubble and orbifold geometries, we analyse associated 
instabilities. Assuming AdS/CFT correspondence, we discuss the effects of this higher 
derivative bulk coupling on a specific matrix model near the critical points of the boundary gauge theory at 
finite temperature. Finally, we propose another phenomenological model on the boundary which mimics 
various phases of the bulk space-time.

\end{quotation}
\vskip 2in
September 2006\\
\end{titlepage}
\vfill
\eject

\section{Introduction}

Black hole in AdS space has a remarkable property that it undergoes Hawking-Page (HP) 
phase transition\cite{H-P}. Asymptotically AdS space allows two kinds of black hole configurations. 
By comparing their sizes with respect to the AdS scale, one can characterise these holes. While 
the small black holes have  horizon sizes less than that of the AdS scale, the big black holes
are larger than the AdS scale. These small black holes, however, are unstable 
with negative specific heat; leaving big black holes as the stable configurations 
in AdS space. Furthermore, it was noticed that, as we tune the temperature 
close
to the inverse of the AdS scale, there is a first order phase transition.
At a temperature below the inverse AdS scale, system prefers thermal AdS space, while
at higher temperature, it is the big black hole phase which minimises the energy of the 
system. This crossover is known as HP transition. Via AdS/CFT correspondence \cite{Maldacena:1997re}, this 
phenomenon was found to have its imprint on the gauge theory residing on the boundary. 
Witten argued \cite{Witten:1998zw} that, on the boundary, the HP transition represents a large 
$N$ deconfinement transition of the gauge theory at strong coupling.

Certain analytical continuation of the black hole metric in AdS space gives bubble of 
nothing solution \cite{BMR,BR}. These are the analogues of Witten's Kaluza-Klein bubbles in flat space-time 
\cite{EW}.
Bubble spacetime corresponds to time dependent configuration  and, as we review later, in 
five dimensions, the boundary metric is  $dS_3 \times S^1$. By using AdS/CFT 
correspondence, one then hopes to learn about gauge theory on time-dependent geometries.
An important ingredient in the AdS/CFT correspondence is the principle of holography 
\cite{Witten:1998qj}. According to this principle, the physics of a gravitational theory is dual to a 
different theory in one lower dimension. Conversely, given a dual theory on a boundary, we 
must consider all possible bulk space-times whose boundaries have the same intrinsic 
geometry 
as the background of the dual theory. Now, given that the boundary metric is $dS_3\times 
S^1$, 
there exists another bulk geometry known as AdS orbifold. These are the five 
dimensional 
analogues of BTZ black holes \cite{BN,BNO}. These orbifolds are however unstable and below a 
critical size of the boundary $S^1$, they decay to bubbles of nothing \cite{RT,BLS}. A bubble, once formed, 
expands exponentially and fills up the whole space-time. Though in AdS, this is similar to the decay of the Kaluza-Klein space to nothing. For other studies involving decay 
into bubbles of nothing, obtained by analytically continuing black hole solutions see \cite{tanayc,dumitrus,dumitrun}.

In this paper, after briefly reviewing the HP transition and orbifold decay in section 2, 
we analyse the response of these phenomena as we perturbatively increase 
the gravitational 
strength . Our study is 
partly motivated by recent works in \cite{RS, BS, AMMPR}. In these papers, authors have argued in different 
ways that a version of HP transition occurs even at weak coupling gauge theory. By AdS/CFT 
dictionary, this would show up as a transition in strongly coupled gravity theory in the 
bulk. Noting the fact that string theory in AdS space is as yet poorly understood, 
we study a much simpler system in this paper. We add higher derivative terms in 
the supergravity action and study their effects on HP transition as well as on orbifold 
decay. We note here that  higher derivative terms would
arise in gravity action due to $\alpha^\prime$ corrections in underlying 
string theory.
While a study with a general class of higher derivative terms would be desirable, 
in this work, we consider only the effects due to Gauss-Bonnet(GB) terms. 
One advantage of working with GB correction to the gravity action is that the black holes, bubbles 
and orbifolds can be constructed explicitly. 

In section 3, we analyse the black holes in GB theory with a particular focus on 
their phase structures in five space-time dimensions. The phase structure depends 
crucially on the GB coupling. For certain range of coupling, there exists three black hole 
phases. We call them, small, intermediate or unstable and big black hole phase. 
It turns out that there are two first order phase  transitions. One of them is  from small 
black hole to  the big one at a temperature scale much lower than that of inverse AdS 
curvature. The other  one is similar to that of usual HP transition where a crossover 
occurs from thermal  AdS to the big black hole phase. We compute the change in HP 
temperature in powers of the GB coupling at the crossover. 

In section 4, we study the bubbles in GB theory. We find that there exist bubble of 
nothing solutions for any value of the asymptotic circle. This is unlike the case in 
simple AdS gravity, where the bubble to exist, the circle size needs to be less than a 
critical value. Consequently, by computing energy densities of the bubble and 
orbifold spacetimes, we argue that orbifolds are always unstable and decay to bubble of 
nothing. The decay rate can then be easily computed by identifying the bounce solution.

Many papers in the recent past have analysed the partition function of free ${\cal{N}} 
=4$ super Yang-Mills theory and argued that the large $N$ deconfinement 
transition occurs 
even at zero coupling \cite{BS, AMMPR}. In fact, it turned out that the transition appears exactly at the 
Hagedorn temperature of the low temperature thermal AdS phase. Subsequently, 
non-perturbative $1/N$ effects near the Hagedorn transition was studied in \cite{HL}. This 
has been analysed in the other limit of the `t Hooft coupling, $\lambda \rightarrow \infty$
by proposing a phenomenological matrix model 
\cite{AGLW,Basu:2005pj,Alvarez-Gaume:2006jg}. 

It is known that GB corrections arise in heterotic string theory, see for example \cite{BD}. In type IIB 
theory the corrections start off with $R^4$.
In the case with Gauss Bonnet terms we do not expect the boundary 
theory on $S^3\times S^1$ to be ${\cal N}=4$ Yang Mills theory. 
However in the limit $\alpha^{'} \rightarrow 0$ the boundary theory 
should reduce to the strongly coupled SYM theory.
With the $R^2$ corrections turned on, the gravity theory should correspond 
to some deformation of ${\cal{N}} =4$ SYM. In Section 5, we study this effective 
theory by using a phenomenological matrix model proposed in \cite{AGLW}. 
This model is characterised by two parameters which we call, following \cite{AGLW}, 
$a$ and $b$. Generally, $(a,b)$ depend on the gauge theory temperature and the
't Hooft coupling $\lambda$. Following AdS/CFT, the effect of adding higher derivative terms 
in the bulk translates to $\lambda$ corrections to the boundary gauge theory.
{\it Assuming} an universal nature of the $(a,b)$ model 
around the critical points, we analyse the $\lambda$ dependence of 
parameters $(a,b)$ around the HP points. We do this numerically in section 
5. 

Finally, in section 6, 
we construct a toy model which captures the whole phase diagram of the bulk. However, this
requires introduction of four parameters in the matrix model potential. 
These four parameters again depend on the temperature as well as the gauge coupling. 
We then study the qualitative behaviour of this model. This paper ends with a discussion of our
results. We hope to report on a  similar analysis for the type IIB theory with $R^4$ term 
in a future publication\cite{DMMS}.

\section{Black hole, bubble and AdS orbifold}

In this section, we briefly recall black holes, bubbles and AdS orbifolds 
in AdS gravity in five dimensions. We review the instability associated with
the AdS orbifolds and calculate the decay rate of these orbifolds to 
bubbles
in the supergravity limit. Furthermore, we compute an enhancement of the decay rate due to 
string wrapping around a compact direction of the AdS orbifold. We also briefly review, 
following a suggestion due to Horowitz \cite{GH}, as to how this decay may be catalysed by tachyon 
condensation in string theory.

\bigskip

\noindent{\bf Black Hole:}

\bigskip

Black hole in pure AdS-gravity is 
parametrised by a single parameter associated with the energy of the hole. Denoting
this parameter as $m$, we may write the metric of the black hole as
\begin{equation}
ds^2 = -(1 + {r^2\over l^2} - {m\over r^2})\, dt^2 + (1 + {r^2\over l^2} -
{m\over r^2})^{-1} dr^2 + r^2 (d\theta^2 + {\rm cos}^2\theta d\Omega_2^2),
\label{schmet}
\end{equation}
where $l$ is related to the cosmological constant present in the
action. This metric asymptotically approaches AdS space. The singularity is at $r = 0$ and 
the horizon is located at $r_+$, where $r_+$ is
a solution of equation
\begin{equation}
1 + {r^2\over {l^2}} - {m\over{r^2}} = 0.
\end{equation}
The Euclidean version of the metric is free of any conical singularity if the Euclidean 
time 
has certain periodicity. Defining 
\begin{equation}
r = r_+ + \Big({r_+\over{2 l^2}} + {m\over{2 r_+^3}}\Big)\rho^2,
\end{equation}
near $r = r_+$, we can write the metric as 
\begin{equation}
ds^2 = d\rho^2 + \Big({r_+\over{ l^2}} + {m\over{ r_+^3}}\Big)^2\rho^2 d\chi^2
+ r_+^2 d\Omega_3^2,
\end{equation}
where, we have used $t = i \chi$. 
From here it follows that the metric is conically non-singular if 
$\chi$ has a period
\begin{equation}
\Delta \chi = { 2 \pi l^2 r_+^3\over{r_+^4 + m l^2}} = {2 \pi r_+ l^2\over{2 r_+^2 + 
l^2}}.
\label{chip}
\end{equation}
Inverse of this periodicity is then identified with the black hole temperature. We note that, 
for a fixed 
temperature (above a certain critical value), we always get two black hole 
solutions with two different horizon sizes. We distinguish these two by 
calling them 
small and big black holes. 
At the critical temperature, both these black holes meet. It turns out that the 
smaller black hole is unstable with negative specific heat while the bigger one is 
stable. The free energy of the black holes  is given by,
\begin{equation}
F = {2 \pi^2 r_+^2\over \kappa_5} \Big( 1 - {r_+^2\over l^2}\Big).
\end{equation}
Here, $\kappa_5$ is related to the five dimensional gravitational constant.
We note that if the size of the black hole is larger than the AdS scale $l$, the free energy 
becomes negative. Since this is less than the free energy of thermal AdS space,  
there is  a first order phase transition. From 
(\ref{chip}), we see that for $r_+ = l$, $T = T_c = {3\over{2\pi l}}$. So the transition 
occurs from thermal AdS space phase to the black hole phase as we increase the temperature 
beyond $T_c$. The crossover between these two geometries is known as HP 
transition \cite{H-P}.

\bigskip
\newpage
\noindent{\bf Bubble:}

\bigskip

The other space-time that is of our interest is the AdS bubble \cite{BMR, BR, RT, 
BLS}. The metric 
can be obtained 
by analytically continuing 
( $t \rightarrow i \chi$ and $\theta \rightarrow \pi/2 + i \tau$) the black hole solution 
given in (\ref{schmet}).
We get
\begin{equation}
ds^2 = (1 + {r^2\over l^2} - {m\over r^2})d\chi^2 + (1 + {r^2\over l^2} 
-{m\over r^2})^{-1} dr^2 + r^2 (-d \tau^2 + {\rm cosh}^2\tau d\Omega_2^2).
\label{bubmet}
\end{equation}
If $\chi$ is restricted to the period as in (\ref{chip}), the metric is non-singular for
$r \ge r_+$. This geometry is known as a bubble of nothing solution. For large $r$, at any 
time $\tau$, the metric is $\chi$ circle times a two sphere. This circle 
collapses at $r = 
r_+$. 
However, the two sphere approaches a finite size $r_+^2 {\rm cosh}^2\tau$. This two sphere is
the boundary of the bubble. We see that the metric is time dependent and is asymptotically  
$dS^3 \times S^1$. As can be seen from the dotted line in Figure 6, below a certain 
critical value of $\Delta \chi$, for a given $\Delta \chi$, there are two possible bubble 
solutions. The smaller one, however, is expected to be unstable as its Euclidean 
continuation
suffers from having modes with negative ${\rm mass}^2$ \cite{MP}. The critical size of the 
$\chi$ circle, above which there are no bubble solutions, is given by
\begin{equation}
\Delta\chi_c =  {l\pi\over \sqrt 2}~~~~{\rm for} ~r_c ={l\over{\sqrt 2}}. 
\label{crit}
\end{equation}

\bigskip

\noindent{\bf AdS orbifold:}

\bigskip 

The AdS orbifold, that we consider here, has been discussed in great detail in \cite{BN, BNO, CAI }.
A five dimensional AdS space is defined as the universal covering space of
a surface which obeys
\begin{equation}
-x_0^2 + x_1^2 + x_2^2 + x_3^2 + x_4^2 - x_5^2 = -l^2,
\end{equation}
where, as before, $l$ is the AdS curvature radius. The orbifold is obtained by simply 
identifying points along the boost
\begin{equation}
\xi = {r_+ \over l} (x_4 \partial_5 + x_5\partial_4),
\end{equation}
where $r_+$ being an arbitrary constant. Since the norm of the boost is given by
$\xi^2 = r_+^2 (-x_4^2 + x_5^2)/l^2$, $\xi^2$ can be positive or negative. However, 
to avoid
closed time-like curves, the region $\xi^2 < 0$ is removed from space-time.
In an appropriate coordinate
system this orbifolded space can be represented as
\begin{equation}
ds^2 = (\tilde r^2 - r_+^2) (-d\tilde t^2 + {l^2\over r_+^2} {\rm
cosh}^2({r_+\tilde t\over l})d\Omega_2^2)
+ {l^2\over{\tilde r^2 -r_+^2}} d\tilde r^2 + \tilde r^2 d\phi^2,
\end{equation}
where $ r_+ \le \tilde r \le \infty$ and $ 0 \le \phi \le 2\pi$.
Further defining
\begin{equation}
\tilde r^2 = r_+^2 (1 + {r^2\over{l^2}}), t = {r_+ \tilde t\over l},\tilde \chi = r_+\phi,
\end{equation}
the metric becomes,
\begin{equation}
ds^2  = (1 + {r^2\over l^2}) d\tilde \chi^2 + ({1 + {r^2\over l^2}})^{-1}dr^2 +
r^2[-dt^2 + {\rm cosh}^2t d\Omega_2^2].
\label{orbif}
\end{equation}
We should note here that, in these coordinates, $0 < r < \infty$. The Euclidean 
version
of this space-time ($t\rightarrow -i\theta -i\pi$) clearly resembles thermal AdS once 
we
reinterpret the periodic coordinate $\tilde \chi$ as Euclidean time. We record the metric 
here for
later use:
\begin{equation}
ds^2   = (1 + {r\over l^2}) d\tilde \chi^2 + ({1 + {r^2\over l^2}})^{-1}dr^2 +
r^2[d\theta^2 + {\rm cos}^2\theta d\Omega_2^2].
\label{euorbif}
\end{equation}

\bigskip

\noindent{\bf Instabilities and decay rates:}

\bigskip

We now see from (\ref{orbif}) and (\ref{bubmet}) that both these
space-times have the same asymptotic geometry. The boundary is time dependent and is 
given by
$dS^3 \times S^1$. However, a notable difference is while the orbifold exists for 
any size of 
the asymptotic $S^1$, the bubble appears only when this boundary circle has a maximum 
critical
size. The size is given by the expression in (\ref{crit}). Boundary energy 
densities of the
orbifold and the bubble are computed in \cite{BR}. They are given by
\begin{equation}
\rho_{\rm orbi} = -{1\over{64\pi Gl}}, ~~\rho_{\rm bubble} = -{1\over{16\pi Gl^3}}(m 
+{l^2\over 4}).
\label{den}
\end{equation}
Let us now consider the case where the size of the boundary circle is less than
the critical value given in (\ref{crit}). In the bulk, we can have both the bubble or 
the
orbifold geometry. However, in view of equation (\ref{den}), we see that the
orbifold will decay to the bubble of nothing by radiating away its energy \cite{RT, BLS}. This is 
the 
analogue of Witten's decay 
of Kaluza-Klein vacuum to a bubble of nothing \cite{EW}\footnote{When embedded in 
supersymmetric theory, one employs antiperiodic boundary condition of the fermions around 
the circle. This breaks supersymmetry completely.}.

As in the case of Kaluza-Klein decay, it is possible to find the bounce 
solution
which mediates this decay. This was discussed in some detail in \cite{BLS}. 
As analysed there, it is the Euclidean
continuation of the smaller Schwarzschild black hole which acts as a 
bounce.
The metric for the bounce is therefore
\begin{equation}
ds^2 = (1 + {r^2\over l^2} - {m\over r^2})d\chi^2 + (1 + {r^2\over l^2} 
-
{m\over r^2})^{-1} dr^2 + r^2 (d\theta^2 + {\rm cos}^2\theta d\Omega_2^2).
\label{bouncemet}
\end{equation}

Having identified the bounce, we can calculate the semiclassical decay rate from the
orbifold to the bubble by evaluating the action difference between
Euclideanised orbifold (\ref{euorbif}) and the bounce (\ref{bouncemet}). Though individual
actions diverge due to large volume, the difference remains finite once we require same
asymptotic boundary conditions for $\tilde\chi$ in (\ref{euorbif}) and for $\chi$ in 
(\ref{bouncemet}).
This is obtained by setting
\begin{equation}
{\sqrt{1 + {R^2\over l^2}}} \beta_{\tilde\chi} = {\sqrt{1 + {R^2\over l^2} - {m\over
R^2}}}\beta_\chi.
\label{bc}
\end{equation}
Here $\beta_{\tilde\chi}$ is the period of $\tilde\chi$ in (\ref{euorbif}) and $\beta_\chi$ 
is the period of $\chi$ in (\ref{bouncemet}); the expression of the later is given in
(\ref{chip}). Now the difference in actions is given by
\begin{eqnarray}
\Delta I &&= I_{\rm bounce} - I_{\rm orbifold} \nonumber\\
         &&\nonumber\\
         &&= {2\times 4\over{16 \pi G l^2}}[ \int_{0}^{\beta_\chi}d\chi\int_{r_+}^Rr^3 dr \int
              d\Omega_3^2 - \int_{0}^{\beta_{\tilde\chi}}d\tilde \chi \int_{0}^R r^3 dr \int 
d\Omega_3^2]\nonumber\\
         &&\nonumber\\
         &&= {\omega_3\over {8G}}({ {r_+^3 l^2 - r_+^5}\over{2 r_+^2 +  l^2}}).
 \label{acdiff}
\end{eqnarray}
Here $\omega_3$ is the volume of unit three sphere. In getting the last expression we have 
made use of the boundary condition (\ref{bc}).

\bigskip

\noindent{\bf Enhancement of decay rate due to string wrapping the circle:}

\bigskip

On generic grounds, we expect an enhancement of
orbifold decay rate when a Nambu-Goto string wraps around the circle $\tilde\chi$.
This is what we intend to compute in this subsection.
We noted that the decay of the orbifold requires a bounce solution with
a negative mode in its spectrum of small fluctuation. The small Euclidean AdS-Schwarzschild 
black
hole that we analysed in the previous section
has such a mode \cite{MP}. In what follows, we will assume that the presence of a
string does not remove this non-conformal negative mode. With this assumption,
it is now easy to see how the decay rate changes as we wrap a string
with action
\begin{equation}
S = T \int d^2\xi {\sqrt{{\rm det} \gamma}}
\end{equation}
along $\tilde\chi$ direction of the orbifold and $\chi$ direction of the bounce solution. 
Here,
$\gamma$ is the
induced metric on the string.
We expect that the change in the decay rate will be proportional to the
exponential of the action difference $\Delta S = S_{\rm bounce} - S_{\rm orbifold}$.
This quantity can be easily computed as follows:
\begin{eqnarray}
\Delta S &&= S_{\rm bounce} - S_{\rm orbifold}\nonumber\\
         &&\nonumber\\
         &&= T \int d^2\xi {\sqrt{{\rm det} \gamma_{\rm bubble}}}
             - T \int d^2\xi {\sqrt{{\rm det} \gamma_{\rm orbifold}}}\nonumber\\
         &&\nonumber\\
         &&= T\int_0^{\beta_\chi} d\chi \int_{r_+}^{R} dr
             - T \int_0^{\beta_{\tilde\chi}} d\bar{\chi} 
               \int_0^{R} dr\nonumber\\
         &&\nonumber\\
          && = T{2\pi l^2 r_+^2\over{2 r_+^2 + l^2}}.
\label{enhan}
\end{eqnarray}
To get to the last step, we have used (\ref{bc}) and also made a large $R$ approximation\footnote{
A similar computation was performed in \cite{ST} in the context of Witten's bubble}.
From the above expression we see that the decay rate increases with $r_+$. This, in turn,
implies that enhancement is larger for larger size bubble in the final state.
We see from the above expression that $\Delta S$ increases with string tension $T$. 
However,
for a string with large energy density, we would have to go beyond test string 
approximation.

\bigskip

\noindent{\bf Decay via tachyon:}

\bigskip

In a recent paper \cite{GH}, Horowitz has argued that a black string can catalyse 
Witten's decay process. When a black brane is wrapped around a compact circle, the circle 
size becomes a function of the position. For a suitable choice of brane, regardless of its asymptotic size,
this circle can reach string scale at the brane horizon. In fact, 
tuning the charges of the branes, this circle size can be made to vary very slowly. 
For an anti-periodic boundary condition of the fermions around this circle, 
one expects a tachyonic mode to appear as the size shrinks to string 
scale. This mode may then
induce a topology changing process by pinching off the circle at the horizon. This, in turn, creates 
a bubble. A concrete example of this is the $D3$ brane in ten dimension. For our purpose, 
instead of a flat space-time, let us consider $N$ $D3$ branes filling up the boost orbifold $R^{1,1}/Z$. In 
the near horizon limit one gets AdS orbifold in five dimensions \cite{BLS}. On the other hand, 
after the tachyon condensation, we have a bubble of the kind that we have been considering. Since this decay 
is catalysed by a string scale process, one would expect the rate to be much faster than the one through 
supergravity bounce.

In the next section we discuss black holes, bubbles and orbifolds in GB theory. We study how the above 
features change as a function of the GB coupling.

\section{Gauss-Bonnet black holes}

We start by considering $(n+1)$ dimensional gravitational action in 
the presence of a negative cosmological constant $\Lambda$ along with a  
GB term. 
\begin{equation}
I = \int d^{n+1} x {\sqrt{-g_{n+1}}}\Big[{R\over{\kappa_{n+1}}} - 2 \Lambda 
+ \alpha (R^2 - 4 R_{ab}R^{ab} + R_{abcd}R^{abcd})\Big].
\label{baction}
\end{equation}
This action possesses  black hole solutions which we call GB black holes \cite{BD, RM, Nojiri:2001aj, CNO, RCAI, CN, TM} . In the above action, $\alpha$ is the GB coupling. As the higher derivative corrections 
are expected to appear from the $\alpha^\prime$ corrections in underlying string theory, we will
often refer to such corrections  as $\alpha^\prime$ corrections in this paper. The 
metric of these holes can be expressed as 
\begin{equation}
ds^2 = - V(r) dt^2 + {dr^2\over{V(r)}} + r^2 d\Omega_{n-1}^2,
\label{bblackmetric}
\end{equation}
where $V(r)$ is given by
\begin{equation}
V(r) = 1 + {r^2\over{2 \hat \alpha}} - {r^2\over{2 \hat \alpha}}\Big[1 - {4 
\hat\alpha\over{l^2} 
}+ {4 \hat\alpha m\over{r^n}}\Big]^{1\over 2}.
\label{bv}
\end{equation}
We first define various parameters that appear in the above equation. $d\Omega_{n-1}^2$
is the metric of a $n-1$ dimensional sphere. $l^2$ is related to the cosmological 
constant as $l^2 = - n(n-1)/(2 \kappa_{n+1} \Lambda)$. 
Furthermore, we have defined $\hat \alpha = (n-2)(n-3) \alpha \kappa_{n+1}$, where 
$\kappa_{n+1}$ is the $n+1$ dimensional gravitational constant. The parameter $m$ in 
(\ref{bv}) is related to the energy of the configuration as
\begin{equation}
M = {(n-1) \omega_{n-1} m\over{\kappa_{n+1}}},
\label{bmm}
\end{equation}
where $\omega_{n-1}$ is the volume of the $n-1$ dimensional unit sphere.
Asymptotically, the metric (\ref{bv}) goes to AdS space, since in this limit
\begin{equation}
V(r) = 1 + \Big[{1\over{2 \hat \alpha}} - {1\over{2 \hat \alpha}}\Big(1 - {4 
\hat\alpha\over{l^2}
}\Big)^{1\over 2}\Big] r^2.
\label{ads}
\end{equation}
We see from here that  the metric is real if and only if
\begin{equation}
\hat \alpha \le l^2/4.
\label{cond}
\end{equation}
In our discussion, we will always consider $\hat \alpha$ satisfying the above bound.
The metric (\ref{bblackmetric}) has a central singularity at $r=0$. 
The zeros of $V(r)$ correspond to the locations of the horizons. 

In five dimension, for which $n=4$, there is a single horizon at
\begin{equation}
r^2 = r_+^2 = {l^2\over 2}\Big[ -1 + {\sqrt{1 + {4 (m - \hat\alpha)\over{l^2}}}}\Big].
\label{bhorizon}
\end{equation}
We note here that for a black hole to exist $m > \hat \alpha$. 

Thermodynamics of these black holes can be obtained via standard Euclidean action calculation.
Such calculations were performed, for example, in \cite{CN}. Following these computations, the free 
energy and 
temperature can be written down as
\begin{eqnarray}
F = &&{\omega_{n-1} r_+^{n-4}\over{\kappa_{n+1} (n-3) (r_+^2 + 2\hat \alpha)}}
\Big[(n-3) r_+^4 (1 - {r_+^2\over{l^2}}) - {6 (n-1) \hat\alpha r_+^4\over {l^2}}\nonumber\\
   && + (n-7) \hat \alpha r_+^2 + 2 (n-1) \hat \alpha^2\Big],\nonumber\\
&&\nonumber\\
T = && {(n-2)\over{4 \pi r_+ (r_+^2 + 2\hat\alpha)}}\Big[r_+^2 + {n-4\over{n-2}}\hat \alpha 
+ {n\over{n-2}} {r_+^4\over l^2}\Big].
\label{ft}
\end{eqnarray}
The black hole entropy is given by 
\begin{equation}
S = \int T^{-1} \Big({\partial M\over{\partial r_+}}\Big)dr_+ = {4\pi \omega_{n-1} 
r_+^{n-1}\over{\kappa_{n+1}}}\Big[1 + {n-1\over{n-3}}{2 \hat \alpha\over r_+^2}\Big],
\end{equation}
and the specific heat is
\begin{eqnarray}
&&C = {\partial M\over{\partial T}}\nonumber\\ 
  &&\nonumber\\
  &&= {4 \pi (n - 1) \omega_{n-1} 
r^{n-3} (r^2 + 2 {\hat \alpha})^2 [ \hat\alpha l^2 (n -4) + r^2 ( l^2 (n -2) +n 
r^2)]\over{ \kappa_{n+1}[\hat \alpha r^2 (6 n r^2  - l^2 (n -8)) + r^4 (n r^2 - (n 
-2) l^2 )
-2(n - 4)\hat\alpha^2 l^2]}}.
\label{spheat}
\end{eqnarray}   
Many interesting features of the GB black holes, related to local and global stabilities, can be 
inferred from a detailed study of the thermodynamic quantities. 
In the rest of the section, we proceed to do so by considering the holes in five dimensions 
($n =4$). Let us first introduce two dimensionless quantities
\begin{equation}
\bar \alpha = {\hat \alpha\over {l^2}}, {\rm and} ~~\bar r = {r_+\over{l}}.
\end{equation}
We would like to express various thermodynamic quantities in terms of these dimensionless 
constants. The free energy given in (\ref{ft}) can be written as
\begin{equation}
F = -{\omega_3 l^2 \over {\kappa_5 (\bar r^2 + 2\bar\alpha)}}\Big[
     \bar r^6 + ( 18 \bar \alpha - 1) \bar r^4 + 3 \bar \alpha \bar r^2 - 6 \bar \alpha^2\Big].
\label{bfree}
\end{equation}
It then follows from (\ref{bfree}), that within the range of allowed value of the coupling 
$\bar \alpha$ (see (\ref{cond})), $F$ starts  
being positive at  $\bar r =0$ and changes sign only once as we increase $\bar r$.
The number of extrema of the free energy, however, crucially depends 
on $\bar \alpha$. In particular, when $\bar \alpha$ is in the region
\begin{equation}
0 < \bar \alpha \le {1\over{36}},
\label{rangeone}
\end{equation}
$F$ has three extrema. At these points, $F$ takes non-zero positive values. 
However, for
\begin{equation}
{1\over{36}} \le \bar \alpha \le {1\over 4},
\label{rangetwo}
\end{equation}
$F$ has no extremum for any non-zero $\bar r$. It starts with a nonzero value at $\bar r =0$, 
then decreases monotonically and becomes 
negative at large $\bar r$. Typical behaviour of the free energy as a function of $\bar r$ is 
shown Figure 1. We will refer back to this plot when we analyse the stability of these holes.
\begin{figure}[ht]
\epsfxsize=8cm
\centerline{\epsfbox{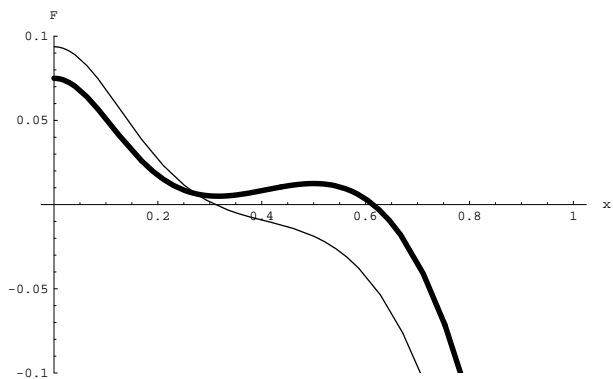}}
\caption{{\small{Free energy as a function of $x = \bar r$ for different values of
$\bar \alpha$. The thicker line is for $\bar \alpha = 1/40$ and the other one 
$\bar \alpha = 1/32$.}}}
\end{figure}

For now, we turn our attention to the temperature of the black holes. It follows from
(\ref{ft}) that the temperature is given by \footnote{
In the limit  $l \rightarrow \infty $ ($ \Lambda = 0 $) 
this solution reduces to the asymptotically flat Gauss Bonnet black hole.
The temperature is then given by 

\begin{equation}
T= \frac{r_+}{2 \pi ( r_+^2 + 2 \hat{\alpha} ) } \nonumber
\end{equation}

For finite value of $ \hat{\alpha} $, temperature begins with zero value at $ 
r_+  = 0$ and gradually increases for small $ r_+ $. Finally 
it reaches a maximum value at
$ r_+ = \sqrt{2 \alpha} $ and then again goes towards zero at large $ r_+ $. 
Since the temperature has a maximum, above this critical value 
there is no black hole solution. 
At any temperature below there are two black hole solutions, small and large. 
The small black hole has positive specific heat and is locally stable. 
The larger one is unstable due to its negative specific heat. This is to be 
contrasted with the Schwarzschild black hole solution without $R^2$ correction, 
where we have only one unstable solution existing at all temperatures. 

}

\begin{equation}
T = {\bar r + 2 \bar r^3\over{2 \pi l (\bar r^2 + 2\bar \alpha )}}.
\label{fivet}
\end{equation}
At $\bar r = 0$, temperature starts out from zero and, regardless of the value of $\bar \alpha$,
it increases for small $\bar r$. 
However, at larger ${\bar r}$, the number of 
extrema depends on $\bar \alpha$. In the region given in (\ref{rangeone}),
there are two of these extrema. Both of these disappear as we increase $\bar \alpha$ 
to region (\ref{rangetwo}). A plot of the temperature as a function of ${\bar r}$ is shown in 
Figure 2.
\begin{figure}[ht]
\epsfxsize=8cm
\centerline{\epsfbox{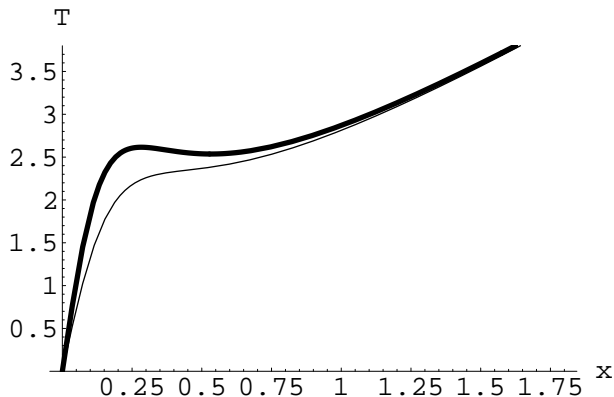}}
\caption{{\small{Temperature as a function of $x = \bar r$ for different values of
$\bar \alpha$. The thicker line is for $\bar \alpha = 1/44$ and the other one
$\bar \alpha = 1/30$.}}}
\end{figure}
\begin{figure}[t]
\begin{center}
\begin{psfrags}
\psfrag{T1}[][]{$T_1$}
\psfrag{T2}[][]{$T_2$}
\psfrag{T3}[][]{$T_3$}
\psfrag{TC}[][]{$T_c$}
\epsfig{file=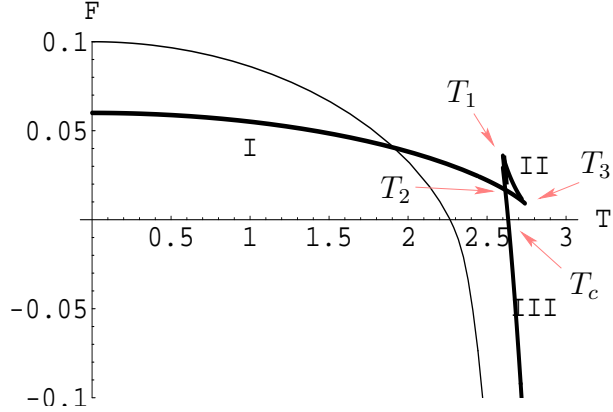, width= 8cm,angle=0}
\end{psfrags}
\vspace{ .1 in }
\caption{{\small{Free energy as a function of temperature. The thicker one is for $\bar \alpha = 
1/50$ while the other one is for $\bar \alpha = 1/30$.}}}
\end{center}
\end{figure}


To examine the phase structure of these black holes, it is instructive to
consider the behaviour of the free energy as a function of temperature (for different values
of $\bar \alpha$). From (\ref{bfree}) and (\ref{fivet}), it is possible to construct the 
temperature dependence of the free energy. However, the analytical expression is not very 
illuminating. Therefore, we plot the nature of the free energy as a function of temperature in 
Figure 3. This plot is for two different values of $\bar \alpha$ belonging to the two different 
regions given in (\ref{rangeone}) and (\ref{rangetwo}). Note that, as we increase $\bar \alpha$
from region (\ref{rangeone}) to region (\ref{rangetwo}), nature of $F$ changes at a critical 
value $\bar \alpha = \bar \alpha_c = 1/36$. We therefore study these two regions separately.

\bigskip

\noindent{\bf Phase structure for $\bar \alpha \le \bar\alpha_c$:}

\bigskip

When $\bar \alpha \le \bar \alpha_c$, the free energy is shown by the 
thicker 
line in Figure 3. At low 
temperature, it has only one branch (shown as branch I in the figure). 
However, when the 
temperature is increased beyond a certain value (which we call $T_1$), 
two new branches appear (II and III). One of these two branches 
(II) meets branch I at a temperature beyond, say $T_3$, and 
they both disappear. On the other hand, branch III continues to decrease rapidly, 
cuts branch 
I at temperature, say $T_2$, and becomes negative at a temperature which we 
will call $T_c$ in the future.
While computing specific heat using (\ref{spheat}), we find that it is positive for 
branch I, and III. Therefore, these phases correspond to stable black holes. They, 
however,
differ in their sizes; branch I represents smaller sized black holes than 
that of branch III. 
Going back now to branch II, we find that the specific heat is negative. 
We, therefore, conclude that branch II represents an unstable phase of the black hole. 

The above picture is similar to that of the  van der Waals gas. In particular, the Gibbs free 
energy of van der Waals gas, for an isotherm, behaves in a similar manner as we vary pressure. 
A thermodynamic equilibrium state is reached by minimising the Gibbs free energy. 
Likewise, in our case, 
equilibrium state would correspond to branch I of the free energy all the way up to 
temperature $T_2$ and then branch III from temperature $T_2$ and above. The free energy 
curve then remains concave as expected for a thermodynamical system. We, however, note that 
since 
there is a discontinuity of $dF/dT$ at $T=T_2$, one has a first order phase transition at 
$T_2$. 
Two black hole phases would differ from each other at this point by a discontinuous change
in their entropies. We will call these as the first Hawking-Page (HP1) transition for reasons that 
will be obvious later.

This phase structure can be nicely described by constructing a Landau 
function around the critical point. By identifying the dimensionless quantity $\bar r$ as 
an order parameter, we can construct a function $\Phi(T, \bar r)$ as\footnote{To construct the 
Landau function, we employ a method similar to the one discussed it \cite{CM, BM}.}

\begin{equation}
\Phi(T, \bar r) = {\omega_3 l^2\over{\kappa_5}}(3 \bar r^4 - 4\pi l T \bar r^3
                    + 3 \bar r^2 - 24 \pi \bar \alpha l T \bar r + 3 \bar \alpha ). 
\label{lfunction}
\end{equation}
At the saddle point of this function, that is when ${\partial \Phi\over {\partial \bar r}} = 
0$, we get back the expression of the temperature given in (\ref{fivet}). If we then 
substitute back the expression of temperature in to (\ref{lfunction}), $\Phi(\bar r)$ 
reduces to 
the free energy given in (\ref{bfree}). As can be seen from Figure 4, for temperature
 $T < T_2$, $\Phi(T, \bar r)$ has only one global minimum. This corresponds to the small black 
hole phase. However, at $T = T_2$, appearance of two degenerate minima 
suggests a coexistence 
of small and big black hole phases. Finally for temperature beyond $T_2$, only the big black 
holes phase remains (as this phase minimizes the Landau function). Clearly, there is a 
discrete change in the order parameter $\bar r$ at $T=T_2$. This is what 
we expect for a first 
order phase transition.
\begin{figure}[ht]
\epsfxsize=8cm
\centerline{\epsfbox{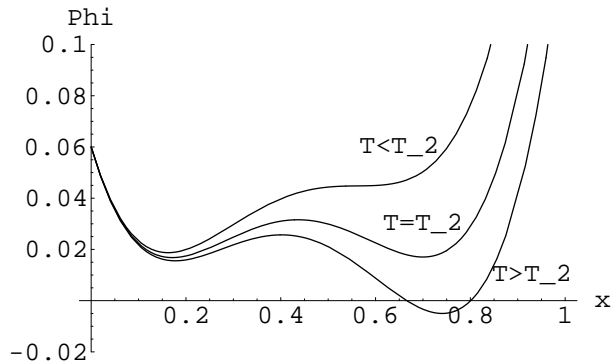}}
\caption{{\small{Landau function $\Phi$ as a function of order parameter x = $\bar r$ for 
different temperatures. We have taken  $\bar \alpha = 1/50$.}}}
\end{figure}

\bigskip

\noindent {\bf Phase structure for $\bar \alpha > \bar \alpha_c$:}

\bigskip

For $\bar \alpha > \bar \alpha_c$, the free energy curve is shown by the thin line 
in Figure 3. Unlike the previous case, free energy and its derivatives do not show 
any discontinuity. Therefore, there is no HP1 transition. Only a single 
black hole phase is found to exist at any temperature.

\bigskip

\noindent {\bf Global phase structure of GB black holes:} 

\bigskip
As discussed earlier, for black holes with $\bar \alpha =0$, there is a crossover from AdS to AdS 
black holes at a critical temperature ${3\over{2\pi l}}$. What happens to this transition as we 
turn on $\bar \alpha$? In this situation, we note that we still have two geometries to 
consider.
First one is again a thermal AdS with metric being the Euclidean continuation of 
(\ref{bblackmetric}). The function $V(r)$ is given in (\ref{ads}). We identify this thermal AdS 
space, having $\bar \alpha$ dependent effective cosmological constant, with zero free 
energy. Now, from Figure 3. we see that above a critical temperature, the free 
energy of the GB black hole becomes 
negative, making it more stable compared to the effective AdS geometry. We identify this as
a HP2 point. This crossover temperature can be computed as a power series in $\bar \alpha$  
and is given by
\begin{equation}
T_c = {3\over{ 2 \pi l}} - {33 \bar \alpha\over{4 \pi l}} + {\cal{O}}(\bar \alpha^2).
\label{Tc}
\end{equation}
We notice here that the GB correction reduces the transition 
temperature. Similar phenomenon was noticed earlier in many AdS-gravity theories with 
higher curvature terms \cite{GUB, GL, KL, CK}. 
The global phase structure is shown in Figure 5.
\begin{figure}[ht]
\epsfxsize=8cm
\centerline{\epsfbox{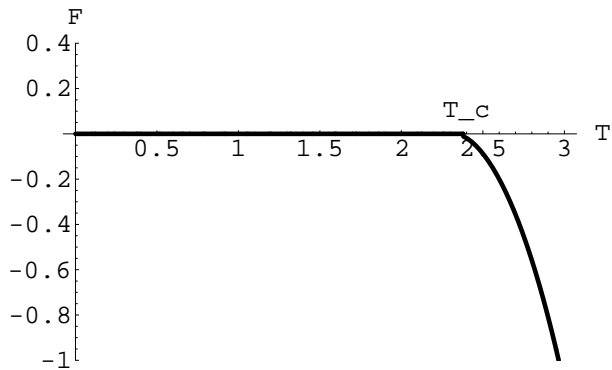}}
\caption{{\small{Global phase structure of GB black holes. For temperature $T<T_c$, AdS 
lowers the free energy, on the other hand for $T >T_c$ the black hole phase is preferred.
This plot is for $\bar \alpha = 1/30$.}}}
\end{figure}

To this end, we would like to point out that the above picture of GB black holes is quite 
similar to that of the five dimensional charged AdS black holes \cite{CEJMO, CEJMT}. 
The stability properties of the charged black holes depend on whether we are considering fixed 
potential ensemble or
fixed charge ensemble. For the case of fixed charge ensemble, various phases of black holes 
resemble that of the thick line in Figure 3. As in our case, small black holes and large black 
holes are separated by a first order phase transition point.
However, a major difference is that for the charged black holes, in fixed charge ensemble, 
thermal AdS is not a solution. Consequently, these holes are globally stable. 
This is unlike GB black holes, where there is a HP2 transition. Below HP2 temperature,
they are unstable.
 
\section{Bubbles in GB theory and instabilities}

In this section, we turn our attention to bubble spacetime in GB theory.
After constructing these bubbles, we compute their energy densities.
We find that due to the presence of these bubbles
with same asymptotic structure as of the AdS orbifolds in GB theory, the orbifolds become 
unstable.
The nature of these decays depend on the value of $\bar \alpha$. In what follows,
we will be mainly interested in studying these instabilities as a function of
$\bar \alpha$. We will also highlight the differences that occur when we compare
the present situation with the one with $\bar \alpha = 0$.

The bubbles in the GB theory can be constructed by analytically continuing the coordinates 
of the GB black holes as $ t \rightarrow i \chi, \theta \rightarrow \pi/2 + i \tau. $ 
Here theta parametrises one of the angles of $d\Omega_{n-1}$ in (\ref{bblackmetric}).
The solutions then takes the form
\begin{equation}
ds^2 = V(r) d\chi^2 + {dr^2\over{V(r)}} - r^2 d\tau^2 + r^2 {\rm cosh}^2\tau d\Omega_{n-3}^2,
\label{bubblemetric}
\end{equation}
where $V(r)$ is given in (\ref{bv}). Above metric is nonsingular in the 
region $r \ge r_+$ if $\chi$ has a periodicity
\begin{equation}
\Delta \chi = {4 \pi \bar r l ( \bar r^2 + 2 \bar \alpha )
\over{(n-2)(\bar r^2 + {n-4\over 
{n-2}}\bar \alpha + {n\over{n-2}} {\bar r^4})}}.
\label{bubbleperiod}
\end{equation}
In the following we will use dimensionless quantity $\Delta\bar \chi = \Delta\chi/l$ 
to parametrise the circle. We first note that at asymptotically large distance, the
metric reduces to $dS_{n-1} \times S^1$ where $S^1$ corresponds to the $\chi$ circle. 
More precisely, up to a conformal scaling by $L^2/r^2$, the boundary metric
becomes
\begin{equation}
ds^2 = d\chi^2 + L^2( - d\tau^2 + {\rm cosh}^2 \tau d\Omega_{n-2}^2).
\label{boundb}
\end{equation}
In the above equation, we have defined
\begin{equation}
L = {\sqrt{2 \hat \alpha}}\Big[ 1 - \Big(1 - {4\hat \alpha\over{l^2}}\Big)^{1\over 
2}\Big]^{-{1\over 2}}.
\label{effl}
\end{equation}
At $r = r_+$, the proper radius of this circle collapses at $V(r) = 0$. However,  
the $n-2$ sphere approaches a finite size $r_+^2 {\rm cosh}^2 \tau$. Therefore  
(\ref{bubblemetric}) represents a bubble of nothing in GB theory with size $r_+$. In the 
rest of this section, we will mostly focus ourselves on the bubbles in five 
dimensions. 

Many of the features of these bubbles can be understood from the behaviour of the 
periodicity $\Delta\bar\chi$ as a function of $\bar \alpha$ and $\bar r$. Firstly, as 
can be seen from Figure 6., for any non-zero $\bar \alpha$, there exists a bubble 
for any size of the $\chi$ circle. This is very much unlike the case when $\bar 
\alpha =0$ where there is a critical radius above which the bubbles are no longer
present. Secondly, for a given $\bar \alpha$ in the range (\ref{rangeone}), and for a 
given periodicity of $\chi$, there can be at most three bubbles of varied sizes.
This can be seen from the solid line in Figure 6. However, as we increase $\bar \alpha$ 
above $1/36$ and go to the range (\ref{rangetwo}), for fixed $\Delta \chi$, we 
get a single bubble of fixed size. This is shown by the dashed line in Figure 6. 
\begin{figure}[ht]
\epsfxsize=8cm
\centerline{\epsfbox{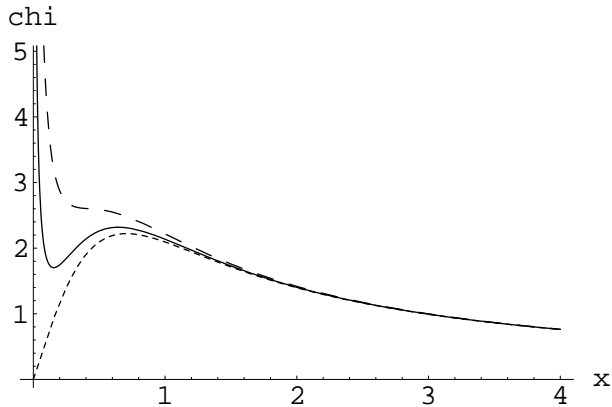}}
\caption{{\small{Plot of $\Delta\bar\chi$ as a function $x = \bar r$. The dashed line
is for $\bar \alpha = 1/34$. The solid line is for $\bar \alpha = 1/50$ and the 
dotted line is for $\bar \alpha = 0$. 
}}}
\end{figure}

Continuing our discussions in five dimensions, let us note that as in the case of 
AdS gravity, in GB theory, we also have another geometry with the same asymptotic 
metric. These are the AdS orbifolds (\ref{orbif}) with 
the AdS curvature $l$ replaced by $L$ as in (\ref{effl}).

The asymptotic boundary of the GB bubble 
spacetime is $dS_3\times S^1$. One would then expect, by AdS/CFT 
correspondence, that some deformation of ${\cal{N}} = 4$ SU(N) Yang-Mills theory at 
large but {\it finite} $\lambda$ should be dual to the bubble. Clearly, 
similar to the AdS bubbles, for the GB bubbles, the CFT lives on a time dependent space. 
The boundary 
stress tensor can be computed from the bulk stress tensor. These bulk stress 
tensor components, for GB bubbles, can easily be obtained and are given by
\begin{eqnarray}
&&T^\tau_\tau = {1\over \kappa_5 L^3}m,\nonumber\\
&&T^\chi_\chi = -{3\over \kappa_5 L^3} m,\nonumber\\
&&T^\theta_\theta = {1\over \kappa_5 L^3}m,\nonumber\\
&&T^\phi_\phi = {1\over \kappa_5 L^3}m,
\label{gbbt}
\end{eqnarray}
where we have parametrised $d\Omega_2$ by the coordinates $\theta$ and $\phi$.
We note here that the components of the stress-tensor computed with respect to 
the orbifold in GB theory\footnote{One easy way to compute these expressions
is to first calculate various components of stress-energy tensor for GB black 
holes. The formalism to compute $T_{\mu\nu}$ for the GB black holes is given
in \cite{PET}. One can then make a proper analytical continuation to get stress tensor for 
the GB  bubbles.}. Positive sign of $T^\tau_\tau$ implies that the
solution has negative mass. 

As of now, we have learned that, in GB theory, bubble space exists for any 
value of $m$. This, in turn, means that we can have a bubble of any size. 
Furthermore, we note that the bubble and AdS orbifold have asymptotically the same 
metric, namely $dS_3 \times S^1$. The energy density of the bubble spacetime
is however less than that of the orbifold. We may, therefore, conclude that
AdS orbifold is an {\it unstable} background in GB theory. It will always
decay to bubble by radiating away its energy. As a result, bubble of any size
will be produced. Once it is produced, due to the time-dependent nature of the 
metric, the radius of the bubble will increase exponentially with time. 

It is easy to compute the decay rate from the orbifold to bubble. This 
can be done, as before, by identifying the bounce solution.

\section{Matrix Model : Some numerical computation}

In the previous sections we have analysed the phase structure of the
gravitational theory in the presence of a higher derivative correction.
In the light of the AdS/CFT correspondence we would now like to analyse
these phases from the gauge theory on the boundary. Specifically,
we will study the thermal aspects of the $SU(N)$ gauge theory on the
boundary
of $AdS_5$ which is $S^3\times S^1$. The effective theory on the boundary
can be described by an unitary matrix model:
\beqa
Z(\lambda,T)=\int dU e^{S_{eff}(U)},
\eeqa
where $U=P\exp(i\int_0^{\beta}A(\tau)d\tau)$ and $A(\tau)$ is the
zero mode of $A_0$ on $S^3$. This is the lightest mode, and the
effective action is obtained by integrating out all the massive modes.
In general ${S_{eff}(U)}$ is a polynomial in the traces of $U$ and its
powers that are allowed by the $Z_N$ symmetry. The coefficients of these
terms depend on the 'tHooft coupling $\lambda$ (that is related
to $\alpha^{'}$ by, $\alpha^{'}\sqrt{2\lambda}=l^2$ from the AdS/CFT
correspondence) and temperature $T$.

These coefficients have been worked out in the weak coupling expansion in
\cite{Aharony:2005bq}. In our analysis we will restrict ourselves to the 
first two terms,
\begin{eqnarray}
S_{eff}(U)=a\tr U \tr U^{\dagger}+\frac {b}{N^2}
(\tr U \tr U^{\dagger})^2,
\end{eqnarray}
where, $a$ and $b$ are functions
of temperature $T$ and $\lambda$.
An order parameter characterising the deconfined phase of the gauge
theory is given by the expectation value of the Polyakov loop
$1/N \expt{\tr U}$.

The saddle point equations for this effective theory are given by,
\begin{eqnarray}\label{saddle}
a\rho+2b\rho^3&=&\rho \mbox{\hspace{0.7in}} 0\le \rho \le \frac{1}{2}
\nonumber\\
&=&\frac{1}{4(1-\rho)} \mbox{\hspace{0.2in}} \frac{1}{2}\le \rho\le 1,
\end{eqnarray}
where $\rho^2=(1/N^2) \tr U \tr U^{\dagger}$. As mentioned this matrix 
model contains only
the first two terms of the effective gauge theory in the weak coupling
expansion. It was shown in \cite{AGLW} that with this truncation, in the 
large $N$ limit one can reproduce
the same thermodynamic
features as of the bulk black hole thermodynamics near the critical 
points.
This model thus appears to fall in the same universality class as that
of the boundary effective gauge theory in the strong coupling limit.

The phase structure in the bulk theory that we have discussed in
Section 3 contains various distinct qualitative features depending on the
value of the correction parameter $\alpha^{'}$. 
For nonzero $\alpha^{'}$,
the phase diagram is modified in the regime where $r_+$ is small compared to
$\sqrt{\alpha^{'}}$. However as long as $r_+$ (the solutions corresponding 
to the black holes at a particular temperature) are greater than 
$\alpha^\prime$, the phase diagram is qualitatively the same as that of the bulk theory without higher derivative corrections. There are two possibilities. 

\begin{itemize}

\item
We can ignore this small black hole solution in the supergravity 
approximation, so that we only concentrate on solutions 
$r_+ > \sqrt{\alpha^{'}}$.
In this domain, it makes sense to compare the bulk physics with that of the 
boundary $(a,b)$ matrix model discussed in the earlier paragraph.

\item
If we include the small black hole solution in the 
supergravity approximation, then in order to reproduce the bulk phases, 
the boundary matrix model needs to be modified. In the next section we 
will propose a matrix model potential that captures the bulk physics including the solution $r_+$ which is less than $\sqrt{\alpha^{'}}$.

\end{itemize}

The following part of this section is devoted to the
study of the $(a,b)$ model numerically,
incorporating the corrections due to the finite 'tHooft coupling
$\lambda$. In this analysis we will take $N \rightarrow \infty$. We
first work in the limit $\lambda \rightarrow \infty$ and then by taking
$\lambda$ large but finite. The main aim is to compute $a$ and $b$ as
functions of $T$ and of $\lambda$ (to the first order in 
$1/\sqrt{\lambda}$).
This will be done by comparing the matrix model potential with the action
on the gravity side with $\alpha{'}$ corrections. In this paper we are
considering the corrections due to the Gauss-Bonnet term.
As mentioned before $R^2$ corrections are not known to occur in the
supergravity limit of type IIB theory. However the following analysis is a
fruitful exercise that can easily be adapted for the $R^4$ terms that 
arise in this theory.

The comparison between matrix model potential and the action of the
bulk theory is valid as long as we can neglect the string loop 
corrections.
The corresponding temperature at which the supergravity description breaks
down is identified as the Gross-Witten transition point in
\cite{AGLW, Alvarez-Gaume:2006jg} on the
matrix model side.

\begin{figure}[t]
\begin{center}
\begin{psfrags}
\psfrag{at}[][]{$a(T)$}
\psfrag{bt}[][]{$b(T)$}
\psfrag{t}[][]{$T$}
\epsfig{file=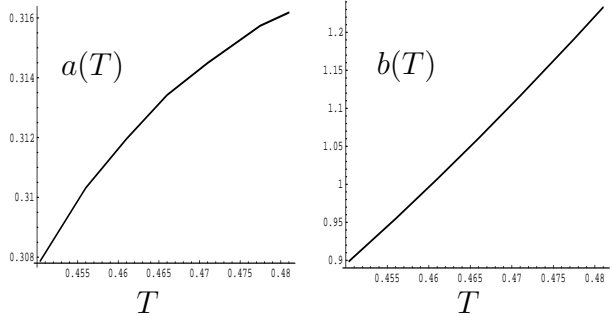, width= 8cm,angle=0}
\end{psfrags}
\vspace{ .1 in }
\caption{Plots of $a(T,0)$ and $b(T,0)$}
\label{atbt}
\end{center}
\end{figure}

Let $T_0$ be the temperature at which the black hole nucleation starts.
For $T>T_0$, it is well known that for the gravity theory without
$\alpha{'}$ corrections, one gets two solutions for the black hole. The
small black hole is unstable and the larger one stable. The larger one
undergoes a Hawking-Page transition at $T=T_c$. It was shown by Witten
\cite{Witten:1998zw} that
thermal $AdS_5$ corresponds to the confined phase of the large $N$ gauge
theory on the boundary. A natural order parameter that characterises the
deconfined phase is the Polyakov loop. In this matrix model it is $\rho$.

Let us study the case without $\alpha{'}$ corrections first.
This corresponds to the $\lambda \rightarrow \infty $ limit. Consider
$T>T_0$, for which we have\footnote{We will set $l$, $\omega_3$, 
$\kappa_5$ to
$1$ in the numerical computations.}

\begin{eqnarray}\label{pot}
2a \rho_{1,2}^2+2b \rho_{1,2}^4+\log(1-\rho_{1,2})+f=-I_{1,2},
\end{eqnarray}
where the $I_{1,2}$ are the actions for the large and small black-holes
respectively and $\rho_{1,2}$ are the corresponding solutions in the
matrix model. The constant $f=\log(2)-1/2$ is added to make the potential
from (\ref{saddle}) continuous at $\rho=1/2$. Since the values of
$\rho_{1,2}$ are
those at the extremum of the left hand side of (\ref{pot}), we have
two more equations that are given by (\ref{saddle}).

\begin{figure}[t]
\begin{center}
\begin{psfrags}
\psfrag{t0}[][]{$T_0$}
\psfrag{tc}[][]{$T_c$}
\epsfig{file=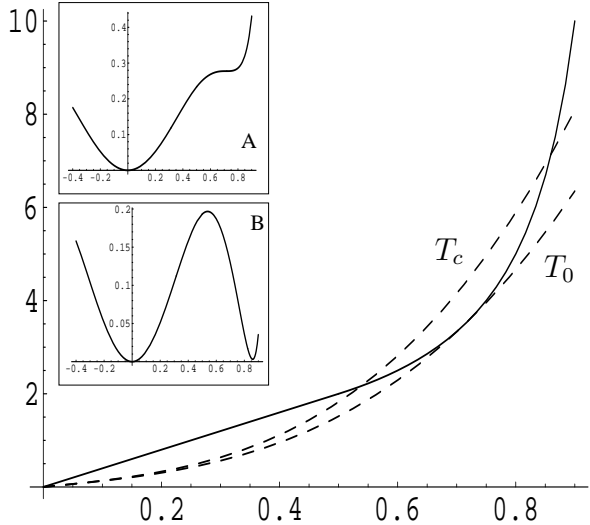, width= 8cm,angle=0}
\end{psfrags}
\vspace{ .1 in }
\caption{The main figure is the plot of left (dashed lines) and
the right hand side (solid line) of
the saddle point equations. $T_0$ is the point where the two roots of
(\ref{saddle}) merge. $T_c$ is the curve corresponding to the Hawking-Page
transition temperature. The inserts A and B are the potentials 
corresponding
to these temperatures}
\label{slope}
\end{center}
\end{figure}

For a given temperature, $T$, $I_{1,2}$ are known from the gravity side,
so the problem now is to
solve the above equations for $a(T)$, $b(T)$ and  $\rho_{1,2}$. We do this
numerically. The solutions are plotted in Figure (\ref{atbt}). Note that
$a(T)$ and $b(T)$ increases monotonically. As a consequence of this
the dashed line in Figure (\ref{slope}) that represents the left hand side
of (\ref{saddle}) moves towards $T_c$ as the temperature is increased thus
generating two solutions for $\rho$ corresponding to the small and
the big black holes in between. $\rho=0$ corresponding to the thermal
$AdS_5$ is always a solution.

\begin{figure}[t]
\begin{center}
\epsfig{file=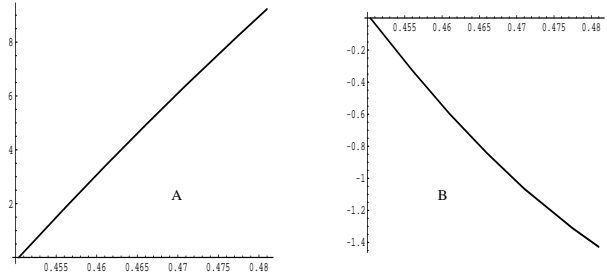, width= 8cm,angle=0}
\vspace{ .1 in }
\caption{Plots of (A) $\partial a(T)/\partial (1/\sqrt{\lambda})$
and (B) $\partial b(T)/\partial (1/\sqrt{\lambda})$}
\label{datbt}
\end{center}
\end{figure}

Having known the variations of $a(T,0)$ and $b(T,0)$ with respect to the
temperature we now incorporate the
$\alpha{'}$ corrections to $I_{1,2}$ to get the first order dependence on
$1/\sqrt{\lambda}$. We have

\begin{eqnarray}
a(T,1/\sqrt{\lambda})&=&a(T,0)+\frac{1}{\sqrt{\lambda}}\frac{\partial 
a(T)}{\partial (1/\sqrt{\lambda})}
\mid_{1/\sqrt{\lambda}=0}+{\cal O}(1/\lambda^{3/2}),\\ \nonumber
b(T,1/\sqrt{\lambda})&=&b(T,0)+\frac{1}{\sqrt{\lambda}}\frac{\partial 
b(T)}{\partial (1/\sqrt{\lambda})}
\mid_{1/\sqrt{\lambda}=0}+{\cal O}(1/\lambda^{3/2}).
\end{eqnarray}

The first order variations of equation (\ref{pot}) with respect to
$1/\sqrt{\lambda}$ gives,

\beqa
2\frac{\partial a(T)}{\partial (1/\sqrt{\lambda})} \rho_{1,2}^2
+2\frac{\partial b(T)}{\partial (1/\sqrt{\lambda})}\rho_{1,2}^4
=-\sqrt{\lambda}\delta I_{1,2}(T).
\eeqa
In the above equations, $\delta I_{1,2}(T)$ are given by,

\beqa
\delta I_{1,2}(T)&=&\alpha^{'}\beta(\delta F_{1,2})\nonumber\\
&=& -\frac{\beta}{\sqrt{2\lambda}}(3r_{1,2}^4+24r_{1,2}^2+9).
\eeqa

Where $F$ is given by eqn(\ref{bfree}). From these we get the values of
$\frac{\partial a(T)}{\partial (1/\sqrt{\lambda})}$ and
$\frac{\partial b(T)}{\partial (1/\sqrt{\lambda})}$ shown in
Figure (\ref{datbt}).

At this point some comments about the sign of $b$ are in order.
In the limit when the t'Hooft coupling $\lambda$ goes to 
infinity numerical computations show that $b$ indeed is positive. However
as we move from $\lambda \rightarrow \infty$ to finite $\lambda$ (that is 
obtained by including $\alpha^{'}$ corrections in the bulk) we see that 
at any particular temperature $\partial b(T)/\partial (1/\sqrt(\lambda))$ 
is negative. Though this numerical calculation shows that $b$ 
decreases from a positive value as we move down towards weak coupling, it 
is not clear whether the sign of $b$ will turn out to be negative or 
positive at weak coupling. In case it turns out to be positive, we 
presume that the results for the gravitational side should correspond 
qualitatively to 
those of the weakly coupled gauge theory. 

The above analysis shows that the behavior of the coefficients as
functions of temperature and $\lambda$ are indeed the ones that
we expect from the phases of the bulk theory as long as we
concentrate on the black hole solutions with $r_{+}>\sqrt{\alpha^{'}}$.
The expansions are carried out about $\lambda \rightarrow \infty$
as it was argued in \cite{AGLW} that the effective theory that is
computed in the weak coupling falls in the same universality
class as the one in the strong coupling limit. The addition of
higher derivative term in the bulk does give information about
$1/\sqrt{\lambda}$ corrections, however this $(a,b)$ model is
unable to capture the phases including the small black hole
solution. In the following section we will analyse this issue,
in detail, by proposing another model which qualitatively reproduces
various bulk phases of section 3.

\section{A modified Matrix model}

In this section we propose a modified (toy) matrix model which incorporates 
some
of the additional qualitative features on the gravity side arised due to 
the GB  term. We find, the minimal
action that would reproduce these features needs to be quartic in $\rho^2$ 
and can be given by
\begin{equation}
S(\rho^2)= 2 [ A_4 \rho^8  -  A_3 \rho^6 +  A_2 \rho^4 + 
(\frac{1-2A_1}{2}) \rho^2 ] \quad ,
\label{action}
\end{equation}
where $A_i$'s are the parameters, which
depend on the temperature as well
as on the coupling constant.
In the limit where the $A_4$ and $A_3$ vanish
we get the $(a,b)$ model
\cite{AGLW}.

The equations of motion ensuing from the action in (\ref{action})
are given as follows.
We write
\begin{equation}
F(\rho) = \frac{\partial S(\rho^2)}{\partial \rho^2} =  [ 8 A_4 \rho^6  -  
6 A_3 \rho^4 +  4A_2 \rho^2 + (1 - 2 A_1) ].
\label{polynomial}\end{equation}
Then the equations in two different regions are
\begin{eqnarray}
\rho F(\rho) &= \rho  \quad, \quad
\quad\quad\quad 0 \leq \rho \leq \frac{1}{2},
\nonumber\\
&= \frac{1}{4(1-\rho)}  \quad, \quad
 \frac{1}{2} \leq \rho \leq 1.
\label{eom}
\end{eqnarray}

The potentials that follows from the above action is given by
\begin{eqnarray}
\!\!\!\!\!\!\!\! V(\rho ) &=& - A_4 \rho^8 + A_3 \rho^6 - A_2 \rho^4  + 
A_1 \rho^2 \quad , \quad\quad 0 \leq
\rho \leq \frac{1}{2} \quad,\nonumber\\
&=& - A_4 \rho^8 + A_3 \rho^6 - A_2 \rho^4  + (A_1-\frac{1}{2}) \rho^2
- \frac{1}{4}\log[2(1-\rho)] + \frac{1}{8} \quad , \quad \frac{1}{2} \leq 
\rho \leq 1 \nonumber\\ 
\end{eqnarray}

Let us analyze the solutions of equations of motion
given by (\ref{eom}).
The fact that there are four parameters instead of two has made the 
analysis technically more involved than
$(a,b)$ model\cite{AGLW}. For various ranges of parameters the model
shows different qualitative behaviour. As we will see we need to impose 
necessary restrictions on the parameters so that the model reproduces 
the features that we found on the gravity side.
Before analyzing the solutions one comment is in order.
In the following we will find the analog of small stable black hole
appearing as a minimum of the potential but it always comes with
an additional maximum of the potential. We do not have
on the bulk side a solution corresponding to this maximum.
We interpret this solution as a possible decay mode of
the small stable black hole which may be due to some
stringy mechanism.

The behaviours of the solutions are encoded in the polynomial $F(\rho)$ given in 
(\ref{polynomial}). We begin with the coefficient of the lowest order term
$A_1$. From (\ref{action}) we see in order to make $\rho = 0$ tachyon-free
we need $0< F(0) \leq 1$ {\it i.e.}
$0 \leq A_1 < 1/2 $. Once that is imposed we consider the next coefficient $A_2$.
As we see on the bulk side our action should admit ( in one phase)
3 solutions that correspond to a small black hole, an intermediate black hole
and a big black hole. A necessary condition for the existence of three solutions
is $ A_2 >0$.
Though we get this constraint from a different argument it agrees with 
\cite{AGLW}. Thus our model at a vanishing limit of higher coefficients
reduces to $(a,b)$ model.

Restrictions on the higher coefficients are slightly more
cumbersome and depends on the positions of the turning points
of the polynomial $F$. In that context it is useful to consider 
the quadratic polynomial in $\rho^2$:
$ f(\rho^2) = (1/\rho)\frac{\partial}{\partial\rho} F(\rho)$.
This is given by $f(x) = 48 A_4 x^2 - 24 A_3 x + 8 A_2$. 
The zeroes of $f$ determine the non-trivial turning
points of $F$. In terms of this polynomial $f$ the 
two different ranges of $\bar\alpha$ correspond to the
following constraints:

$\bullet$ $\bar\alpha > 1/36$: 
For $f(1) = 48 A_4 - 24 A_3 + 8 A_2  < 0$ 
there is one turning point at some $0 < \rho_- < 1 $. 
With parameters
in this range we can have either two solutions (one maximum and one minimum
of potential) or no solution.
There is no way we can obtain three solutions in this phase.
Moreover, from the restriction on the parameters it is clear that
the range of parameters is not continuously connected with the 
corresponding range where $(a,b)$-model is valid ({\it i.e. $A_4 = A_3 = 0$}).
We identify this phase with the range of $\bar\alpha$
which corresponds to  $\bar\alpha > 1/36$ on the gravity side. 
However, that is not sufficient to ensure that there is
always one minimum that correspond to the single black
hole on the bulk side. For that we need to impose a 
further restriction on the coefficients such that,
the turning point satisfies $\rho_- < 1/2$ and 
$F(\rho_-) > 0$. Then we always get a maximum for
$\rho < 1/2$ ( that is in the region with no cut)
and a minimum of the potential. As the parameter
varies the position of this minimum changes from 
the $\rho < 1/2$ region
to the $\rho > 1/2$ region.
So this phase corresponds to the restrictions:
$f(1) = 48 A_4 - 24 A_3 + 8 A_2  < 0$, 
$\rho_- < 1/2$ and $F(\rho_-) > 0$.

$\bullet$ $\bar\alpha < 1/36$:
Again we look for turning points in the range $0 < \rho \leq 1$.
For $f(1) = 48 A_4 - 24 A_3 + 8 A_2  > 0$
either we get two turning points which we call $\rho_-$
and $\rho_+$  ($\rho_- < \rho_+$) or none of them. 
That gives rise to 3 possibilities: the number of solutions
could be 4 (consists of two maxima and two minima), 
or 2 (consists of one maximum and one minimum)or 0.
This phase is continuously connected with that of the $(a,b)$
model and we identify this phase with the range of $\bar\alpha$ 
given by $\bar\alpha < 1/36$ on the gravity side.
The more detailed structure of the solutions depends on 
the position of the turning points $\rho_-$
and $\rho_+$
and the values of the polynomial $F(\rho)$ at the turning points.
Let us first consider $F(\rho_-) > 0$ with $ 0 < \rho_- < 1/2$.
There are two possibilities: (i) If $F(1/2) < 0$ 
we have two solutions, one maximum and other minimum
in $\rho < 1/2$ range. The minimum corresponds to the
stable small black hole. We may or may not have two more solutions
in the range $\rho > 1/2$. If we have two solutions they
would correspond to intermediate and big black hole.
(ii) If $F(1/2) > 0$ 
we have one solution (maximum) in $\rho < 1/2$ and the other 
(minimum) in $\rho > 1/2$. This minimum corresponds to
the big black hole.
The remaining possibilities are (iii) $\rho_- < 1/2$, $F(\rho_-) < 0$
and (iv) $\rho_- > 1/2$, $F(1/2) < 0$. In both of these cases
there is no solution in the range $0 < \rho < 1/2$. 
Finally if there is no turning point and $F(1/2) < 0$
there will be no solution in the range $0 < \rho < 1/2$.
Since the analysis on the bulk side then requires 
that there is a solution for $1/2 < \rho < 1$
we need to impose the following constraint, namely,
there should exist
some value of $\rho$, $ \rho_- < \rho_0 < 1$
such that $4\rho_0(1-\rho_0)F(\rho_0) > 1$.
That will give one maximum and one minimum
in the range $1/2 < \rho < 1$ that  corresponds
to the small and the big black hole.

Thus we see for both the phases we need additional restrictions
which shows there are 
regions of parameters that does not agree with
the features of gravity phase. This suggests the fact that 
in the strongly coupled gauge theory there
are restrictions on various parameters. It may be 
interesting to calculate these parameters from field theory
set up (for weakly coupled gauge theory) and compare
the values with the restrictions obtained above.

\begin{figure}[ht]
\epsfxsize=13cm
\centerline{\epsfbox{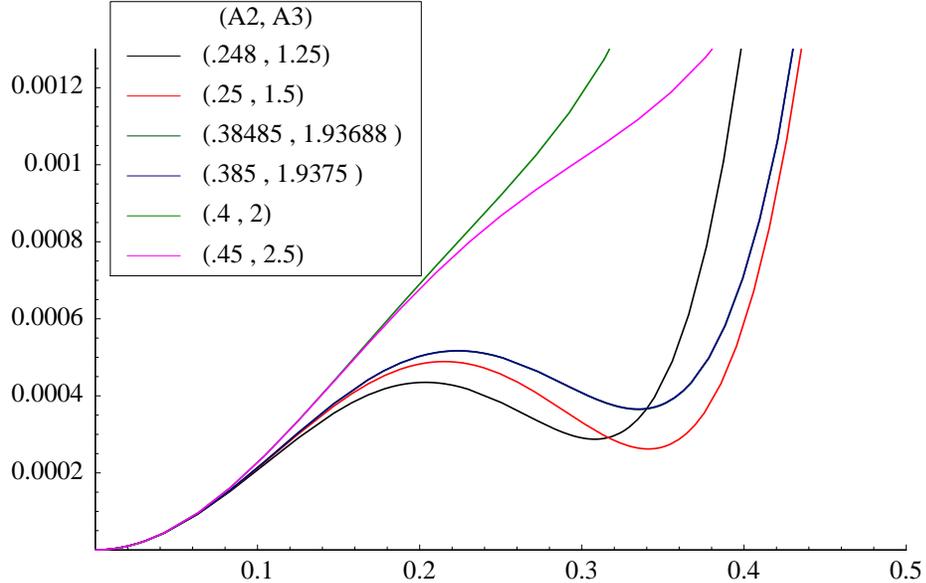}}
\caption{{\small{Potential as function of $\rho$ for the range $0\leq\rho
\leq 1/2 $ for increasing values of $A_2$ and $A_3$. The different values
of ($A_2$, $A_3$) are given above. The plots associated with (.38485,
1.93688)and (.385, 1.9375) are not distinct in this scale.}}}
\label{potential1}\end{figure}

\begin{figure}[ht]
\epsfxsize=13cm
\centerline{\epsfbox{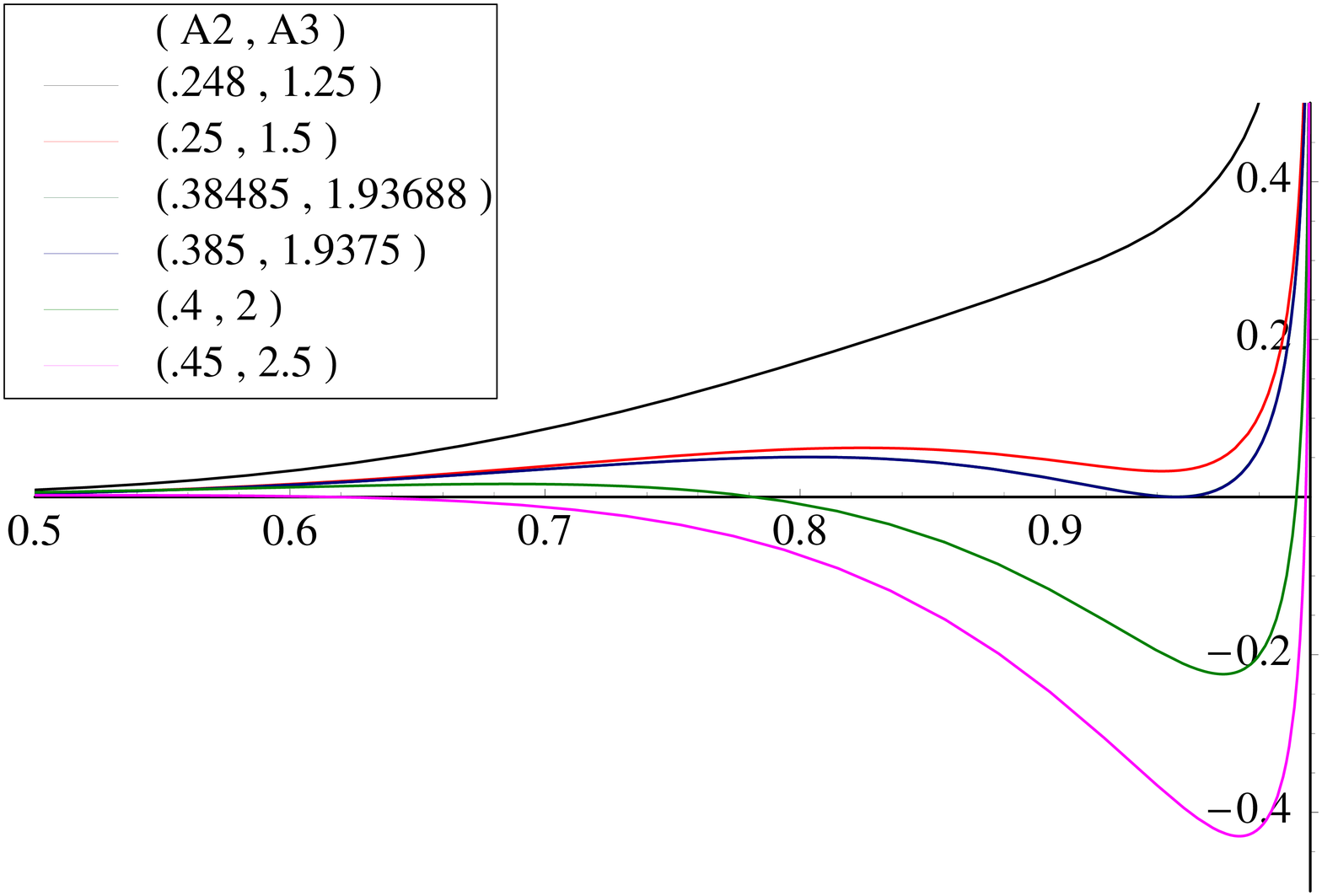}}
\caption{{\small{Potential as function of $\rho$  for the range $1/2 
\leq\rho \leq 1$ for increasing values of $A_2$ and $A_3$. The values of
($A_2$, $A_3$) used in the plots are given above.  The plots associated 
with (.38485 , 1.93688) and (.385 , 1.9375) are not distinct in this
scale.}}}
\label{potential2}\end{figure}

\begin{figure}[t]
\begin{center}
\begin{psfrags}
\psfrag{I}[][]{I}
\psfrag{II}[][]{II}
\psfrag{III}[][]{III}
\psfrag{IV}[][]{IV}
\psfrag{C}[][]{C}
\epsfig{file=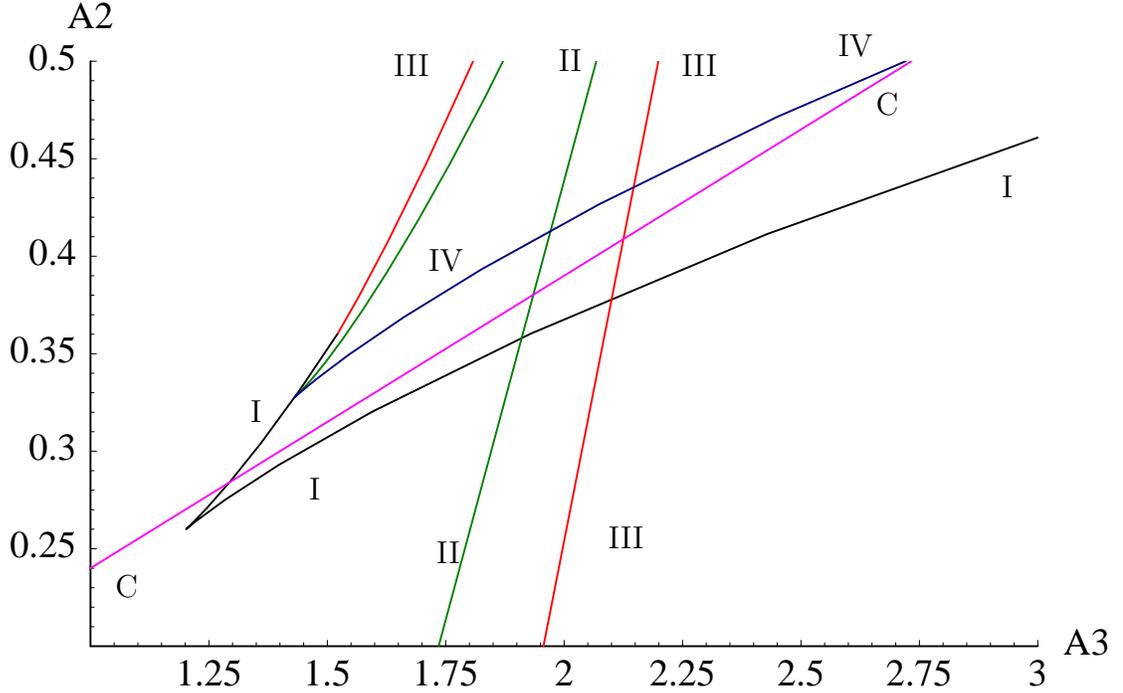, width= 15cm,angle=0}
\end{psfrags}
\vspace{ .1 in }
\caption{{\small{Parametric plots of different critical points in the 
$A_2$-$A_3$ plane. We  choose $A_1=0.25$ and $A_4=2.083$.
In the region which is below or on the left side of curve I the saddle 
points $\rho_\pm$ cease to exist. In the region above curve IV the
potential of $\rho_+$ vanishes. Curve II corresponds to HP transition and 
on curve III the saddle points
$\rho_1$ , $\rho_2$ merge.}}}
\label{transition}
\end{center}
\end{figure}

In order to discuss the variation of potential with
parameters it is useful to give a graphical description.
We have four parameters, so for the sake of graphical
description we restrict number of parameters to 2.
We consider only the phases that corresponds
to the $\bar\alpha < 1/36$. We take 
fixed values of $A_1$ and $A_4$ and
study the features with the variation of two other parameters. We have 
chosen the values to be $A_1=.025$ and $A_4=2.083$.
We have given plot of the potential against $\rho$ in Fig.\ref{potential1}
and \ref{potential2}. In order to make the extrema explicit we choose 
different scales for the
potential for two different ranges of $\rho$, namely, $0 \leq \rho\leq 
1/2$  and
$1/2 \leq \rho\leq 1$. The values of $A_2$ and $A_3$ are
decreasing from the curve in bottom to that in top 
in  Fig.\ref{potential1} and from the curve on top to that in bottom
in Fig.\ref{potential2}. There is always one minimum at
$\rho=0$ where the potential vanishes.
As we will see in
Fig.\ref{transition} we need to choose values of $A_2$ and $A_3$ 
restricted within a particular region outside which the features
that we get from the bulk will be absent.
In the following, we give $V(\rho)$ vs. $\rho$ plots for different values 
of $A_2$ and $A_3$:

\begin{itemize}

\item $( A_2, A_3) = (.45, 2)$: Here we get two solutions: 
one maximum and one local minimum in the range  
$\rho\leq 1/2$ ({\it i.e} where there is no cut) 
and no solution at $\rho\geq 1/2$. We identify the minimum
with the small stable black hole. This corresponds to
low temperature behaviour of GB black hole where we get only one small 
black hole solution.

\item $( A_2, A_3) = (.4, 2)$: Here we get four solutions: 
In addition to the above maximum and minimum in the
range $\rho\leq 1/2$ we get one more local maximum and 
one more 
local minimum appearing in the range $1/2 \leq \rho\leq 1$. 
These latter maximum and minimum can be identified with
the intermediate black hole and the big black hole.
We identify this with the
nucleation of the big black hole and 
intermediate unstable black hole in 
the gravity picture.

\item $( A_2, A_3) = (.385, 1.9375)$: For further decrease of the 
parameters, the heights of the local minimum in the range
 $0\geq\rho\geq 1/2$ (Fig. 
\ref{potential1})increases and the height of the local minimum
in the range $1/2 \geq \rho \geq 1$ decreases(Fig. 
\ref{potential2}) .
At this value of the parameters the heights of the two minima 
become equal. We can 
identify
this point with a transition from small black hole to big black hole on 
the gravity side which is termed as HP1 transition.

\item $( A_2, A_3) = (.38485, 1.93688)$: (Due to close proximity this plot 
appears on the top of the earlier plot and not distinguishable in
the present scale.) The height of the minimum in the range $\rho\geq 1/2$ 
becomes zero and thus equal to the potential at $\rho=0$. On the
gravity side this corresponds to energy of big black hole reaching zero 
and becoming equal to that of thermal AdS triggering HP2 transition.

\item  $( A_2, A_3) = (.25, 1.5)$: Here we get two solutions
because in the region $\rho\leq 1/2$ the
local minimum and local maximum is on the verge of disappearing. 
However, the two solutions
in the range $\rho\geq 
1/2$ will remain with the height of the minimum in  $\rho\geq 
1/2$ keeps on decreasing. 
This corresponds to the point beyond which the 
small black hole on the gravity side disappears.

\item $( A_2, A_3) = (.248, 1.25)$: As we decrease $A_2$ and $A_3$ 
further, the solutions in the range $\rho\leq 1/2$ 
cease to exist (Fig.\ref{potential1}). The minimum in the 
range $\rho\geq 1/2$ (Fig. \ref{potential2}) 
becomes more and more deeper. This is in keeping with the 
fact that, at high
temperature on the gravity side the only stable
configuration remains is the big black hole.
\end{itemize}

As we see from the above analysis the coefficients decrease with 
temperature, unlike the behaviour of the coefficients in the $(a, 
b)$-model.
This can be interpreted as the temperature gradient of the coefficients at 
the first order of inverse 't Hooft coupling has a negative sign
relative to that at the zeroeth order. At this range, where appreciable 
$1/\lambda$ correction is taken into account, the contribution at first
order dominates over that at zeroeth order.

Here we give a graphical presentation of the behaviour of
the solutions using a parametric plot of 
different critical points in the
$A_2$-$A_3$ plane in Fig.\ref{transition} keeping  $A_1$ and $A_4$ fixed
as above. 
As we vary the parameters we
encounter a curve IV in the $A_2$-$A_3$ plane, above which the saddle point 
associated with the small black
hole has energy negative. From the analysis of black holes
on the gravity side, it follows that the small black hole energy is always
greater than thermal AdS ensuring the stability of the latter. So, in what 
follows, we restrict ourselves to
the region below curve IV.

In the region bounded by IV, III and I, there are three saddle points. One 
is $\rho = 0$ which corresponds to
the thermal AdS. There are two more saddle points: a local maximum at 
$\rho=\rho_1$ and a local minimum at $\rho
= \rho_2$. The latter corresponds to the small black hole that we obtain 
on the gravity side. There is no
solution analogous to $\rho_1$ in the gravity side. In the region bounded 
by II, III, IV and I, there appears
two more saddle points. One of them $\rho=\rho_3$ is a local maximum and
the other one $\rho=\rho_4$ is a local minimum. They correspond to the 
intermediate, and the
stable big black hole respectively.
In the region on the left hand side of curve I,
the saddle points $\rho = \rho_1 , \rho_2$ cease to exist.
In the region above the curve IV, as we have already mentioned, the 
potential
of the saddle point $\rho=\rho_1$ becomes negative showing the energy of 
the associated small black hole on the gravity side becomes less than
that of thermal AdS.

Similarly, the thermal history  (for this choice of parameter)
can be obtained from Fig.\ref{transition} 
as follows. As we mentioned earlier,
$A_2$ and $A_3$ will decrease with temperature along the curve C. As we 
follow the curve C from right to left,
we find the $\rho=0,\rho_1 , \rho_2 $ are the solutions on the right of curve III.
As we cross curve III, we encounter two additional saddle points 
$\rho=\rho_3,\rho_4$.
Crossing the curve II corresponds to the Hawking-Page transition.
As we cross curve I, the saddle point corresponds to the small black hole 
disappears.

Like the general case, here also as we see in the  region bounded by the 
curve I, along with a local 
minimum ( at $\rho = \rho_1$) we
always obtain a local maximum ( at $\rho = \rho_2$ ). We interpret 
this maximum, as we said earlier, as 
a bounce solution through which
the small stable black hole decays. It will be interesting to understand this 
instability on the gravity side.

\section{Discussion}

In this paper, we have discussed phase transition of
asymptotically AdS black hole solutions in presence of
Gauss-Bonnet term.
As long as $\bar\alpha$, strength of the coupling to GB term remains above 
certain critical value
$\bar\alpha_c$,
one gets a single black hole phase at any temperature. However,
as the coupling comes down below the critical value
two additional black holes appear. We called them small and
intermediate black holes. The intermediate black
hole is found to have negative specific heat.
It turns out that this small stable black hole is a local minimum
below a critical temperature. Beyond this temperature small black hole
disappears. We have studied the
associated phase diagram and find that the phase structure resembles that 
of van der Wall's gas.  In
addition to the the standard Hawking-Page transition, we have identified 
one more phase transition where the
two branches of the phase diagram meet. We find the specific heat diverges 
at this new critical point.

From a different perspective, the Euclidean version of the  black hole 
solution has been interpreted as the
bounce mediating the decay of AdS orbifold into the bubbles of nothing. It 
was found earlier that for the
radius of circle above a critical value the AdS orbifold is stable while 
below that bound the AdS orbifold
decays. After adding the higher derivative terms in the action, we find 
that there is a bubble solution that
exists for all values of
the radius and has an energy density lower than that of AdS orbifold. This 
suggests that the AdS orbifold is
unstable for any radius.
One of the stringy feature that we have addressed in this paper is the 
modification of
decay rate in presence of a background string wrapped on the circle and we 
find
the decay rate gets enhanced. It would be interesting to find tachyonic 
decay ensuing
from wrapped black brane in this set up.

Five dimensional theory of gravity usually corresponds to some gauge 
theory on the boundary and the analysis
on the gravity side has natural implications about the gauge theory.
In absence of Gauss-Bonnet term, the gravity theory on Euclidean AdS ( 
along with $S^5$) is known to be dual
to be
pure N=4 SYM on a three sphere at finite temperature and the phase diagram 
associated with the gravity theory
captures thermal history of N=4 SYM on $S^3$. In a similar spirit, we 
expect, dual of this five dimensional
gravity theory in presence of Gauss-Bonnet term is some deformation of the 
above gauge theory
and the phase diagram captures its thermal history. In \cite{AGLW}, 
qualitative features of $N=4$ SYM on
$S^3$ was studied from the perspective of a matrix model. This model is 
phenomenological in nature and is
charecterised by two parameters $(a,b)$. On generic ground, one expects 
these parameters to be
$\lambda$ and $T$ dependent. Appealing to the universal nature of this 
model near the critical points,
we find out $\lambda$ dependence of $(a,b)$. This is done by mapping the 
bulk $\alpha^\prime$ correction to
the boundary.
This method can easily be used to find similar $1/\lambda$ dependence of 
matrix model coefficients in the
case of other higher derivative corrections of the gravity action, such 
as, $R^4$ term in IIB theory.

We have also proposed a modified matrix model that captures the 
qualitative features of the phase diagram of
the bulk theory.
Unlike $(a,b)$ matrix model this model is non-universal and
the phase diagram is
reproduced only in a selected region of the parameter space. In addition 
the temperature dependence of the
coefficients turn out to be different from usual linear increasing 
function. We also find there is a bounce
through which the small black hole can decay.
It will be interesting to identify this
instability on the gravity side.  We hope to
return with some of
these issues in future.
\vspace{.2in}



\noindent {\bf Acknowledgements: }

We have benefited from discussions with Nabyendu 
Das, JianXin Lu, Balram Rai, Shibaji 
Roy, Gautam Tripathi and Spenta Wadia.




\begin{thebibliography}{99}
\newcommand{\np}{{\it Nucl. Phys.} {\bf B}}
\newcommand{\pl}{{\it Phys. Lett.} {\bf B}}
\newcommand{\prd}{{\it Phys. Rev. }{\bf D}}
\newcommand{\prl}{{\it Phys. Rev. Lett.}}
\newcommand{\mpl}{{\it Mod. Phys. Lett. }{\bf A}}
\newcommand{\ijmp}{{\it Int. J. Mod. Phys.}{\bf A}}
\newcommand{\cqg}{\it Class. Quant. Grav.}
\newcommand{\jmp}{\it J. Math Phys.}
\newcommand{\cmp}{\it Comm. Math. Phys.}
\newcommand{\atmp}{\it Adv. Theor. Math. Phys.}
\newcommand{\jhep}{\it JHEP}

\bibitem{H-P} S. Hawking and D. Page, Commun. Math. Phys. 87, (1983) 577.

\bibitem{Maldacena:1997re}
  J.~M.~Maldacena,
  Adv.\ Theor.\ Math.\ Phys.\  {\bf 2}, 231 (1998)
  [Int.\ J.\ Theor.\ Phys.\  {\bf 38}, 1113 (1999)]
 hep-th/9711200.

\bibitem{Witten:1998zw}
  E.~Witten,
  Adv.\ Theor.\ Math.\ Phys.\  {\bf 2}, 505 (1998)
hep-th/9803131.

\bibitem{BMR} D. Birmingham, M. Rinaldi, Phys. Lett. b544 (2002) 316.

\bibitem{BR} V. Balasubramanian and S. Ross, Phys. Rev. D66, (2002) 086002,
hep-th/0205290.

\bibitem{EW} E. Witten, Nucl. Phys. B195 (1982) 481.


\bibitem{Witten:1998qj}
  E.~Witten,
  Adv.\ Theor.\ Math.\ Phys.\  {\bf 2}, 253 (1998)
  hep-th/9802150.

\bibitem{BN} M. Banados, Phys. Rev. D57, (1998) 1068, gr-qc/9703040.

\bibitem{BNO} M. Banados, A. Gomberoff and C. Martinez, Class. Quant. Grav. 15 (1998) 3757, 
hep-th/9805087.


\bibitem{RT} S. Ross and G. Titchener, JHEP 0502 (2005) 021, hep-th/0411128.

\bibitem{BLS} V. Balasubramanian, K. Larjo and J. Simon, Class. Quant. Grav. 22 (2005) 4149, 
hep-th/0502111.

\bibitem{tanayc}
  A.~Biswas, T.~K.~Dey and S.~Mukherji,
  Phys.\ Lett.\ B {\bf 613} (2005) 208
 hep-th/0412124.


\bibitem{dumitrus}
  D.~Astefanesei and G.~C.~Jones,
  JHEP {\bf 0506} (2005) 037
  hep-th/0502162.


\bibitem{dumitrun}
  D.~Astefanesei, R.~B.~Mann and C.~Stelea,
  JHEP {\bf 0601} (2006) 043
hep-th/0508162.


\bibitem{RS} S. Kalyana Rama and B. Sathiapalan, Mod. Phys. Lett. A13 
(1998) 3137, hep-th/9810069.

\bibitem{BS} B. Sundborg, Nucl. Phys. B573 (2000) 349, hep-th/9908001.

\bibitem{AMMPR} O. Aharony, J. Marsano, S. Minwalla, K. Papadodimas and M. Van Raamsdonk,
Adv. Theor. Math. Phys. 8 (2004) 603, hep-th/0310285.

\bibitem{HL} H. Liu, hep-th/0408001.

\bibitem{AGLW} L. Alvarez-Gaume, C. Gomez, H. Liu and S. Wadia, Phys. Rev. D71 (2005) 124023,
hep-th/0502227.

\bibitem{BD} D. Boulware and S. Deser, Phys. Rev. Lett. 55 (1985) 2656.

\bibitem{DMMS} T.K. Dey, S. Mukherji, S. Mukhopadhyay and S. Sarkar, to appear.

\bibitem{GH} G. Horowitz, JHEP 0508 (2005) 091, hep-th/0506166.


\bibitem{CAI} R-G. Cai, Phys. Lett. B544, (2002) 176, hep-th/0206223.

\bibitem{MP} M. Perry, "Instabilities in gravity and supergravity", In {\it Superspace and 
supergravity: Proceedings of the Nuffield Workshop}, Cambridge University Press, 1981.

\bibitem{ST} K. Selivanov and T. Tomaras, JHEP 0210 (2002) 065, hep-th/0207172.

\bibitem{RM} R. Myers, Nucl. Phys. B289 (1987) 701.

\bibitem{Nojiri:2001aj}
  S.~Nojiri and S.~D.~Odintsov,
   Phys.\ Lett.\ B {\bf 521} (2001) 87
  [Erratum-ibid.\ B {\bf 542} (2002) 301]
  arXiv:hep-th/0109122.

\bibitem{RCAI} R-G. Cai, Phys. Rev. D65 (2002) 084014, hep-th/0109133.

\bibitem{CNO} M. Cvetic, S. Nojiri and S.D. Odintsov, Nucl. Phys. B628 (2002) 295.

\bibitem{CN} Y. Cho and I. Neupane, Phys. Rev. D66 (2002) 024044, hep-th/0202140. 

\bibitem{TM} T. Torii and H. Maeda, Phys. Rev. D71 (2005) 124002, hep-th/0504127.

\bibitem{CM} L. Cappiello and W. Muck, Phys. Lett. B522 (2001) 425, hep-th/0107238.

\bibitem{BM} A. Biswas and S. Mukherji, Phys. Lett. B578 (2004) 425, hep-th/0310238.

\bibitem{GUB} S. Gubser, I. Klebanov and A. Tseytlin, Nucl. Phys. B534 (1998) 202, 
hep-th/9805156.

\bibitem{GL} Y. Gao and M. Li, Nucl. Phys. B551 (1999) 229, hep-th/9811019.

\bibitem{KL} K. Landsteiner, Mod. Phys. Lett. A14 (1999) 379, hep-th/9901143.

\bibitem{CK} M. Caldarelli and D. Klemm, Nucl. Phys. B555 (1999) 157, hep-th/9903078.

\bibitem{CEJMO} A. Chamblin, R. Emparan, C. Johnson, R. Myers, Phys. Rev. D60 (1999) 064018,
hep-th/9902170.

\bibitem{CEJMT} A. Chamblin, R. Emparan, C. Johnson, R. Myers, Phys. Rev. D60 (1999) 104026,
hep-th/9904197.

\bibitem{PET} A. Petrov, Class.Quant.Grav. 22 (2005) L83, gr-qc/0504058.

\bibitem{Aharony:2005bq}
  O.~Aharony, J.~Marsano, S.~Minwalla, K.~Papadodimas and M.~Van
Raamsdonk,
Phys.\ Rev.\ D {\bf 71} (2005) 125018, hep-th/0502149.

\bibitem{Basu:2005pj}
  P.~Basu and S.~R.~Wadia,
  Phys.\ Rev.\ D {\bf 73} (2006) 045022, hep-th/0506203.

\bibitem{Alvarez-Gaume:2006jg}
  L.~Alvarez-Gaume, P.~Basu, M.~Marino and S.~R.~Wadia, hep-th/0605041.




\end{thebibliography}
\end{document}